\newif\ifsubmode
\submodetrue

\newif\ifprintfig
\printfigtrue

\ifsubmode
  \documentclass[12pt,preprint]{aastex}
  \received{}
  \revised{}
  \accepted{}
\else
  \documentclass{aastex}
  \usepackage{emulateapj5,apjfonts}
\fi

\newcommand{\SFRD}{\mbox{$\dot{\rho}_{\rm SFR}$}}
\newcommand{\SFRDz}{\mbox{$\dot{\rho}_{\rm SFR}(z)$}}
\newcommand{\tausf}{\mbox{$\tau_{\rm SF}$}}
\newcommand{\fesc}{\mbox{$f_{\rm esc}$}}

\begin{document}

\title{Lyman-Break Galaxies and the Reionization of the Intergalactic Medium\altaffilmark{1}}

% USE FULL NAME 

\author{Henry C. Ferguson, Mark Dickinson\altaffilmark{2}}

\affil{Space Telescope Science Institute, 3700 San Martin Drive,
Baltimore, MD 21218.  ferguson@stsci.edu, med@stsci.edu}

\and

\author{Casey Papovich}

\affil{Steward Observatory, University of Arizona, 933 North Cherry Avenue,
Tucson, AZ 85721.  papovich@as.arizona.edu}

\altaffiltext{1}{Based on observations taken with the NASA/ESA Hubble
Space Telescope, which is operated by the Association of Universities
for Research in Astronomy, Inc.\ (AURA) under NASA contract
NAS5--26555}   
\altaffiltext{2}{Visiting Astronomer, Kitt Peak National
Observatory,  National Optical Astronomy Observatories, which is
operated by the  Association of Universities for Research in
Astronomy, Inc.  (AURA)  under cooperative agreement with the National
Science Foundation.}

%%%%%%%%%%%%%%%
% Abstract -- on a separate page in submit mode
%%%%%%%%%%%%%%%

\ifsubmode \clearpage \fi

\begin{abstract}

Near-infrared observations of Lyman-break galaxies at redshifts $z \sim
3$ are beginning to provide constraints on ages, star-formation
histories, dust content, metallicities, and stellar masses.  At
present, uncertainties of more than an order of magnitude are typical
for many of these parameters.  It is nonetheless interesting to ask what
the stellar-population models imply for the existence and luminosities
of Lyman-break galaxies at higher redshift. To this end we examine the inferred
star-formation rates in two well-studied samples
of galaxies as a function of redshift out to $z = 10$ for various
best-fit and limiting cases.

Taken at face value, the generally young ages (typically $10^{8 \pm
0.5} \,{\rm yr})$ of the $z = 3$ Lyman break galaxies imply that their stars
were not present much beyond $z=4.$ By $z = 6$ the cosmic
star-formation rate $\SFRD$ from the progenitors of these galaxies is
less than 10\% of $\SFRD$ at $z=3 \pm 0.5$, even for maximally-old
models, provided the derivative of the star-formation rate $SFR(t)$ is monotonic. The
escaping Lyman-continuum radiation from such galaxies would be
insufficient to reionize the IGM. Thus other sources of
ionizing photons (e.g.  very massive population III stars) may be
needed, and the more normal Lyman-break galaxies may be a phenomenon
confined to redshifts $z \lesssim 4$. This conclusion changes if
$SFR(t)$ was episodic, and we examine the
parameters of such bursty star-formation that might be consistent with
both the $z=2-4$ luminosity functions and the $z \sim 3$ spectral
energy distributions.

\end{abstract}

%%%%%%%%%%%%%%%
% Keywords
%%%%%%%%%%%%%%%

\keywords{galaxies: high-redshift --- 
	  galaxies: stellar content ---
	  galaxies: formation --- 
	  galaxies: evolution}

\ifsubmode \clearpage \fi

%%%%%%%%%%%%%%%
% Beginning of main text
%%%%%%%%%%%%%%%
 
\section{INTRODUCTION} 
\label{sec:intro}

Studies over the past several years have revealed a nascent population
of galaxies at redshifts $z=2-4$ with properties that are in many 
ways similar to those of local starburst galaxies \citep{GSM96,SGPDA96}.
These galaxies are identified by their strong UV continuum emission
and by the presence of strong spectral breaks at Lyman $\alpha$ and
the Lyman limit (rest-frame 1216{\AA} and 912{\AA}, respectively).
Near-infrared photometry of such Lyman-break galaxies (LBGs) provides
access to the rest-frame optical portion of their spectra, and hence
greatly improves the constraints on their stellar populations
\citep{SY98}.
Two recent studies (\citealp{PDF01}, hereafter referred to as PDF01; and \citealp{SSADGP01}) have explored a broad
range of stellar population models for LBGs, varying
age, star-formation timescales $\tausf$, metallicities $Z$, 
and reddening $E(B-V)$. Our goal in this paper
is to examine the implications of these stellar-population models
for LBGs at higher redshift. This is a simple
thought experiment. In reality we expect star-formation histories
to be more complex than these simple models (which, for example, 
ignore chemical evolution entirely) and we expect galaxy merging to be
extremely important at these high redshifts. Nevertheless, 
the spectral-energy distributions (SEDs) of the individual galaxies at 
$z \sim 3$ should reflect the products of this evolution, and some
of the broad implications for star-formation rates vs. time are 
relatively insensitive to the details.

One motivation for exploring galaxy evolution at $z > 5$ is to
understand the connection between galaxies and the physical conditions
in the intergalactic medium (IGM). Observations of QSO absorption lines
indicate that the IGM was highly ionized out to redshifts $z \sim 6$,
while very recent observations suggest that it was more neutral
at higher redshifts \citep{Betal01,DCSM01}.  It is not known if the
sources of ionization were stars or quasars, or whether the stars
responsible for the reionization had a mass function at all similar to
that observed in the Milky Way.  
%It is also unclear whether
%reionization itself had a major role in regulating subsequent galaxy
%formation, for example by suppressing star formation in low-mass galaxy
%halos \citep{Babul Rees 1992}.  
Estimates
for the number of ionizing photons needed to reionize the IGM range
from 1 to 15 photons per H atom \citep{MHR99,HAM01},
requiring UV luminosity densities at $z \gtrsim 6$ at least as
high as those observed at $z=3$.  If the stellar
populations responsible for reionization formed with an initial
mass function (IMF) similar to that observed in present-day
stellar populations, then the remnants of these populations must 
account for a portion of the light emitted by $z = 3$ LBGs. 
It is thus interesting to explore
whether simple models can provide sufficient reionizing photons at 
$z \sim 6$ without violating the stellar-population constraints
for the $z \sim 3$ LBGs.

% The ages of LBG galaxies at $z \sim 3$ are typically 
% $10^{8 \pm 0.5} \,$yr and are very poorly constrained on a
% galaxy-by-galaxy basis \cite{PDF01; Shapley 2001}.
% However, provided that the star-formation
% history $SFR(t)$ was monotonic (i.e. the star-formation rate
% did not have multiple peaks), the overall stellar masses
% of the LBGs are relatively robust. 
% Age, $\tausf$ and extinction uncertainties
% largely cancel. Therefore, constant star-formation models  with
% maximal masses provide a useful limiting case. For such models
% we find that (1) most of the progenitors of $z = 3$ LBGs would
% have been born at $z < 4.5$, and (2) the UV luminosity density
% at $z = 6$ would have been insufficient to reionize the IGM.
% This finding is mildly inconsistent with the observed luminosity
% function of $3.5 < z < 4.5$ LBGs (Steidel 1999), suggesting either
% that monotonic $SFR(t)$ is not an appropriate model or that the IMF
% is deficient in intermediate-mass stars. 

We review the LBG
stellar population constraints in \S\ref{sec:review}. In 
\S\ref{sec:backintime} we turn the clock back on the stellar-population
models and compute $\SFRDz $ to higher redshift. In \S\ref{sec:implications}
we discuss the implications and in \S\ref{sec:burstsimf} we discuss
the modifications of $SFR(t)$ or the IMF that might be 
required to account for both the $z \approx 4$ LBG luminosity function and reionization.
Throughout this paper we adopt the cosmological parameters
$h,\Omega_{\rm tot}, \Omega_m, \Omega_\Lambda = 0.7,1.0,0.3,0.7.$

\section{Lyman-break galaxy stellar populations}\label{sec:review}

\citet{PDF01} studied a sample of spectroscopically-confirmed LBGs
from the Hubble Deep Field North (HDF) in the redshift range $2.0 \lesssim z
\lesssim 3.5$. The data included UV-optical photometry from WFPC2,
J and H-band photometry from NICMOS, and 
$\rm K_s$-band photometry from the KPNO
4m Mayall telescope \citep{Dickinson98p219}. Fluxes were determined from
profile-weighted photometry, which accounts for the PSF variations and
image blending. 
Stellar-population models from the 2000 version of the \citet{BC93} code
were fit to 31 galaxies, varying metallicity, e-folding timescale
$\tausf$, age, IMF (Salpeter, Miller-Scalo, Scalo), extinction, and
extinction law (\citealp{CABKKS00}, SMC). The geometric mean of
the best-fit ages for the sample is 0.12 Gyr for the solar metallicity
case.  Thus a typical galaxy observed at $z = 3.0$ would have
``formed'' at $z=3.15$. \cite{PDF01} showed there to be very few 
galaxies at $z=3$, even considering those that might have escaped
Lyman-break selection, with colors consistent with significantly 
older ages.

\citet{SSADGP01} analyzed
$G, {\cal R}, J,$ and $K_s$ photometry for a sample of %81
galaxies with spectroscopic redshifts $2.2 < z < 3.4$. Colors were
determined from isophotal apertures on PSF-matched images. Solar-metallicity
stellar-population models from the 1996 incarnation of the 
\citet{BC93} code were fit,
with various values of $\tausf$, age, and extinction.
The \citet{Calzetti97} attenuation law was adopted, and the 
published paper reports results only for the best-fit continuous star-formation
models ($\tausf = \infty$) to the 74 galaxies for which acceptable fits were obtained.
The median best-fit age for this sample is 0.32 Gyr, implying
a formation redshift $z=3.4$ for a typical galaxy observed at $z=3$.

Clearly, the inferred ages for {\it monotonic} star-formation histories
in these two studies are very young.  \citet{PDF01} also found that a
substantial fraction of the stellar mass could be hidden in a
``maximally old'' passively evolving population that formed
instantaneously at $z = \infty$. Inferred LBG masses typically increase
by a factor of 3 in such models. 
For our purposes, the interesting point is that   
the star-formation rate appears unlikely to have been constant
over a Hubble time --- i.e. 
$M_s/t_H$, the stellar mass divided by the Hubble time at the LBG redshift, 
is typically 
much less than the measured SFR at $z \sim 3$. This generic
conclusion is unlikely to be very sensitive to the details of the
stellar-population models.

\section{Turning back the clock}\label{sec:backintime}

In exploring the implications of these models, we shall consider three
limiting cases:  ($i$) a single burst of star formation, ($ii$)
continuous star formation starting at some time $t$, and ($iii$) a
two-burst model. Multiple burst models would be
intermediate between these cases.
Figure 1 shows the star-formation rate as a function
of redshift inferred for each galaxy in the PDF01
sample for solar-metallicity models. 
Models with $0.2 Z_\odot$ give younger
ages and higher star-formation rates.
The top two panels of Figure 1 show single-burst models
with $SFR \propto e^{-t/\tausf}$ and 
continuous star-formation
models.  In the burst models
only one out of the 31 galaxies would have been present at $z = 6$. In
the oldest continuous-star-formation models, six out of 31 or 19\%
would have been present at $z=6$.  Figure 1c shows the
results for the best-fit solar-metallicity continuous-star-formation
models of \citet{SSADGP01}. The models imply that only 17\% of the
galaxies were present at $z=6$. 

Models of type ($iii$) with two distinct episodes of star formation
allow more star formation at higher redshift. \citet{PDF01} fit
maximally-old models to their LBG sample, deriving constraints
on the mass of an old population that formed with a Salpeter IMF
in an instantaneous burst at $z = \infty.$ This model quantifies
how much stellar mass can be hidden ``underneath the glare'' of the younger
population. The star-formation rate predicted at $z=6$ from
such maximally old components is zero, because all star-formation
happened at higher redshift. 
Starbursts induced by mergers are likely to be spread out over a
range of redshift. If the older burst in the LBGs is put at redshift lower 
than $z = \infty$, the mass in the burst must be lower. Rather
than fit a whole suite of models of different burst redshifts, we
can, to a good approximation, scale the allowable mass in
the old component by a power-law fading model. 
By fitting the B-band luminosity vs.
time for $10^7 < t < 2 \times 10^9 \,\rm yr$, we find
$L_{\rm B} \propto t^{-0.8}$ for a Salpeter IMF
for an instantaneous burst in the Bruzual \& Charlot solar-metallicity
models. The B band is chosen because the older burst population
contributes mostly longward of $\lambda_{\rm rest} = 3000$ {\AA} 
(See PDF01, Fig. 19). This fading exponent is slightly 
shallower than that adopted by 
\citet{HP97}, because of the narrower age range used
for our estimate. A Scalo IMF would
fade more gradually, as would a lower-metallicity model.
The allowed mass in a burst as a function of age is
$M(z) = M_{\rm max}({\rm age}/t_H)^{0.8},$ 
where $M_{\rm max}$ is the maximum mass allowed 
in an instantaneous burst formed at $z=\infty$. 
If each galaxy had an instantaneous probability $P(z)$ of forming 
stars in a burst of typical duration $\delta t$, then 
the average SFR per galaxy from an ensemble of such galaxies would be 
$M(z)P(z)/\delta t.$ 
For simplicity we adopt a constant $P(z)$ from $z=10$ to the
observed LBG redshift $z_{\rm obs}$.
(We consider varying $P(z)$ in \S\ref{sec:burstsimf}.)
The ensemble-average star-formation rate is thus 
$\xi(z) = M(z)/(t_{\rm obs}-t_{10})$, where $t_{10}$ is the age
of the universe at $z=10$, and $t_{\rm obs}$ is the age of the universe
at the redshift of the LBG. 
In the current generation of semi-analytic hierarchical models,
the rate of star-formation
due to mergers decreases at $z>3$ \citep{CLBF00,SPF01}. 
Therefore our assumption of constant $P(z)$ puts a higher
proportion of star-formation at high redshift. 

Figure 1d shows the SFR vs. redshift implied by such a stochastic model
for two individual galaxies in the PDF01 sample.
The low-redshift spikes in the star-formation rate correspond to the
young component that dominates the light at the observed redshift; the
star-formation progressing to higher redshift represents the mean for
an ensemble of stochastic bursts. Obviously any single galaxy would
simply show two spikes of star formation for this kind of model, but if
we consider such a galaxy as a proxy for millions of others, the
star-formation history shown in the figure represents the maximal rate
of star-formation due to stochastic bursts as a function of redshift.

The results become clearer if we
consider the entire sample of galaxies.  Figure 2 shows the evolution
of $\SFRDz$ with time relative to that $z = 3$ computed by summing up
the models shown in the previous figures. The top panel shows the
monotonic star-formation histories.  For these cases the inferred
co-moving density of star formation declines dramatically from $z=3$ to
higher redshift.  The stochastic burst model is shown by the solid
curve in Fig. 2b. Even if we put the maximum mass allowed in
stochastic-starbursts at redshifts $z > z_{\rm observed}$, the
star-formation rate at $z=6$ is still a factor of 3 below that at
$z=3$.

\section{Implications}\label{sec:implications}

\subsection{The Luminosity Function at $z \sim 4$ }

All of the star-formation histories considered so far imply a dramatic
decline in star formation rate by $z=4$.  However the observed LBG
rest-frame UV luminosity functions are very similar at $z = 3$ and
$z=4$, and the integrated star-formation rates derived therefrom differ
only by a factor of $1.1 \pm 0.4$ \citep{SAGDP99}. Thus the
star-formation {\it histories} derived from the $z=3$ LBGs are in
direct conflict with the star-formation {\it rates} derived for the
$z=4$ LBGs. 

\subsection{Reionization}

If all of the
ionizing photons come from star formation, \citet{MHR99}
estimate that the amount of star-formation needed is
\begin{equation}
\SFRD \approx 0.013 {\fesc}^{-1}
(\frac {1+z} {6})^3
(\frac {\Omega_b h_{50}^2} {0.08})^2 C_{30}
\,M_\odot \, {\rm yr^{-1} Mpc^{-3}},
\end{equation}
where $\fesc$ is the mean fraction of Lyman-continuum radiation 
that escapes from galaxies, $\Omega_b$ is the baryon density,
$h_{50}$ is the Hubble constant in units of $50\,\rm km \, s^{-1} Mpc^{-1}$,
and $C_{30} = 30 \langle n_{\rm HII}^2 \rangle/\bar {n}_{\rm HII}^2$ 
is the ionized hydrogen clumping factor.
Adopting $\fesc = 0.1$,
the required density of star-formation for reionization 
in this model is
a factor of 1.3 times higher than the dust-corrected 
$\SFRD$ at $z \sim 3$ measured by \citet{SAGDP99}.
\footnote{
The value of $\fesc$ is highly uncertain. Measurements
by \citet{SPA01} give a flux ratio 
$F(900{\AA}/F(1500{\AA}) \approx 0.2$ for a sample of galaxies
$z \approx 3.4$ (but see \citet{GCDF02}).
This is higher than most stellar population
models predict even if $\fesc = 1$. On the other hand, the 
estimated mean dust attenuation factor for Lyman break
galaxies is 4.4 at 
$\lambda = 1500\,${\AA} \citep{SAGDP99},
implying $\fesc \lesssim 0.2$ even
ignoring neutral hydrogen opacity.}
In contrast, the star-formation rates inferred from the 
SED fits imply a sharp decrease in $\SFRD$ between $z=3$
and $z=6$. For the monotonic star-formation histories, this
decrease is at least one order of magnitude. Even
for the case of stochastic bursts the star-formation rate 
is still well below that needed for reionization.
The problem becomes even more severe
if a significant fraction of the baryons are
already collapsed into minihalos at the time of reionization.
In this case the required number of ionizing photons increases by a 
factor of 10-20 \citep{HAM01}, and all models fall
short even if $\fesc = 1$.

\subsection{What Kind of Starbursts are Needed?}\label{sec:burstsimf}

In the discussion above, we adopted a uniform starburst
probability $P(z)$ and found that such a model was unable to produce
enough photons at $z \gtrsim 6$ to account for reionization (while at
the same time fitting $z=3$ LBG SEDs). One simple modification would be
to increase the burst probability at high redshift. Keeping $P(z)$
uniform, we require that bursts occur with uniform probability over the
redshift range $z_{\rm min} < z < z_{\rm max}$ and vary $z_{\rm min}$
and $z_{\rm max}$ until the SFR at $z=6$ equals that at $z=3$.
Independent of $z_{\rm max}$ we find that values of $z_{\rm min} > 4.4$
are required to achieve this.  Thus, LBG evolution
would be characterized by an early epoch of star-formation responsible
for reionization, followed by a lull, followed by increased star
formation at $z \sim 3.$ This kind of behavior might be caused by
reheating of the IGM during reionization \citep{CM01}.
However, such a scenario would increase the discrepancy at $z=4.$

More star formation can be hidden in bursts if the bursts fade faster.
For a first-order estimate, we adopt a power-law fading model $L(t)
\propto t^{-\zeta}$. For a Salpeter IMF in the B band $\zeta = 0.8$.
We vary $\zeta$ until the UV luminosity density at $z=6$ equals that at
$z=3$. We find that a fading exponent $\zeta = -1.1$ is required.  As
shown by the dashed curve in Fig. 2b, such a model still falls short of
the observed luminosity-density at $z=4$, but is within the
uncertainties. If the IMF is a powerlaw $\phi(M)dM \propto M^{-(1+x)}$,
a fading exponent $\zeta = -1.1$ requires an IMF slope $x = 0.5$
compared to the Salpeter value $x = 1.35$ (for an instantaneous-burst
solar-metallicity stellar population). A steeper fading slope $\zeta =
-1.2$ (corresponding to an IMF slope $x = 0.3$) is needed to bring
$\SFRD$ at $z=4$ to within a factor of 1.3 of that at $z=3$. Lower
metallicities require even more top-heavy IMFs.  Options other than
varying the IMF are of course possible (e.g.  evolved stellar
populations could be hidden by dust that builds up over timescales of
$10^8$ to $10^9$ yrs).  However, the requirement for
faster-than-Salpeter fading is robust.  Furthermore, the fading must be
even faster if galaxies on average have more than two burst episodes.

\section{Conclusion}\label{sec:conclusions}

In summary, we find that the monotonic star-formation histories that
best match $z = 3$ LBG spectra fail (by a large factor) to provide
enough photons to reproduce the luminosity density at $z=4$ or to reionize
the IGM at $z \gtrsim 6$.  Even stochastic-burst models, which permit
factors of $3-10$ more mass to be formed at higher redshift, fail to
resolve the shortfall.  We are left with a variety of more complex
alternatives.

(1) If we require that the stellar populations responsible for reionization
formed with typical Galactic IMF ($x \sim 1.35$), and that
such star formation did not show a pronounced gap between
$z=6$ and $z=3$, then we must conclude that the remnants of the
stellar populations responsible for reionization {\it do not
reside in $z=3$ Lyman-break galaxies}. This is possible, for example,
if undetected dwarf galaxies with
number-densities higher than the extrapolation of the LBG luminosity
function  dominate the ionizing background.

(2) The spectral energy distributions of $z=3$ LBGs allow for a
separate epoch of normal-IMF star formation at very high redshift
provided that such star formation ceased by $z \approx 6$, leaving
a gap in star-formation until $z \lesssim 4$. This solution to
the reionization problem glosses over the need to explain the
$z=4$ LBG luminosity function.

(3) Reionization could have been caused by stellar populations
heavily weighted toward massive stars \citep{Larson98,ABN00,ONMW01}.  If this phenomenon was confined to 
high redshift (e.g. high-mass, zero-metallicity Population III
stars), then the remnants could reside in lower redshift LBGs as
black holes or neutron stars. This solution to the reionization
problem also fails to solve the $z=4$ LBG problem.

(4) Both problems can be resolved if the star formation in LBGs
was episodic and the stars formed with a top-heavy IMF.
Bursts of star formation associated with mergers are a natural
consequence of hierarchical models of galaxy formation, and are
incorporated to varying degrees into many of the current semi-analytical models
\citep{KH00,CLBF00,SPF01}. 
With the assumption that the starburst probability $P(z)$ is
constant over $3 \lesssim z < 10$, we find that an IMF slope $x \sim
0.3-0.5$ would be required to explain both the relative constancy of
the LBG luminosity function over the range $2 < z < 4.5$ and
plausibly provide enough star formation at $z \gtrsim 6$ to reionize
the IGM. Top-heavy IMFs could in
principle result from higher ISM pressure during 
mergers (\citealp{PNJ97,CBPT98}; but see \citealp{SVCP98}).
Local tests are difficult to carry out because the remnants of
the massive stars responsible for producing the UV photons at high
redshift are neutron stars or black holes today. In the Galactic
bulge, the best fit slope for the mass-function for 
$M < 1 M_\odot$ 
is $x = 0.33$ \citep{ZCFGORRS00}.
Micro-lensing experiments
\citep{Uetal94,Aetal97} do not yet rule out the possibility
that this slope could have continued up to $100 M_\odot$.
More direct constraints on the star-formation histories of LBGs will 
improve greatly over the next few years with the advent of the
Space Infrared Telescope Facility and the Advanced Camera for Surveys
and ultimately the Next Generation Space Telescope.

\acknowledgements
We would like to thank our collaborators on the HDF observations for
their many contributions to this work. We thank Jennifer Lotz and
Mauro Giavalisco for valuable discussions. Support for this work was
provided by NASA through grant GO07817.01-96A from the Space Telescope
Science Institute, which is operated by the Association of
Universities for Research in Astronomy, under NASA contract
NAS5-26555.

%%%%%%%%%%%%%%%
% Reference List
%%%%%%%%%%%%%%%

%\ifsubmode \clearpage \fi

\bibliographystyle{apj}
\bibliography{apjmnemonic,bib}

%\ifsubmode \clearpage \fi

\ifsubmode \clearpage 

\begin{figure}
\centerline{
\epsscale{0.7}
\plotone{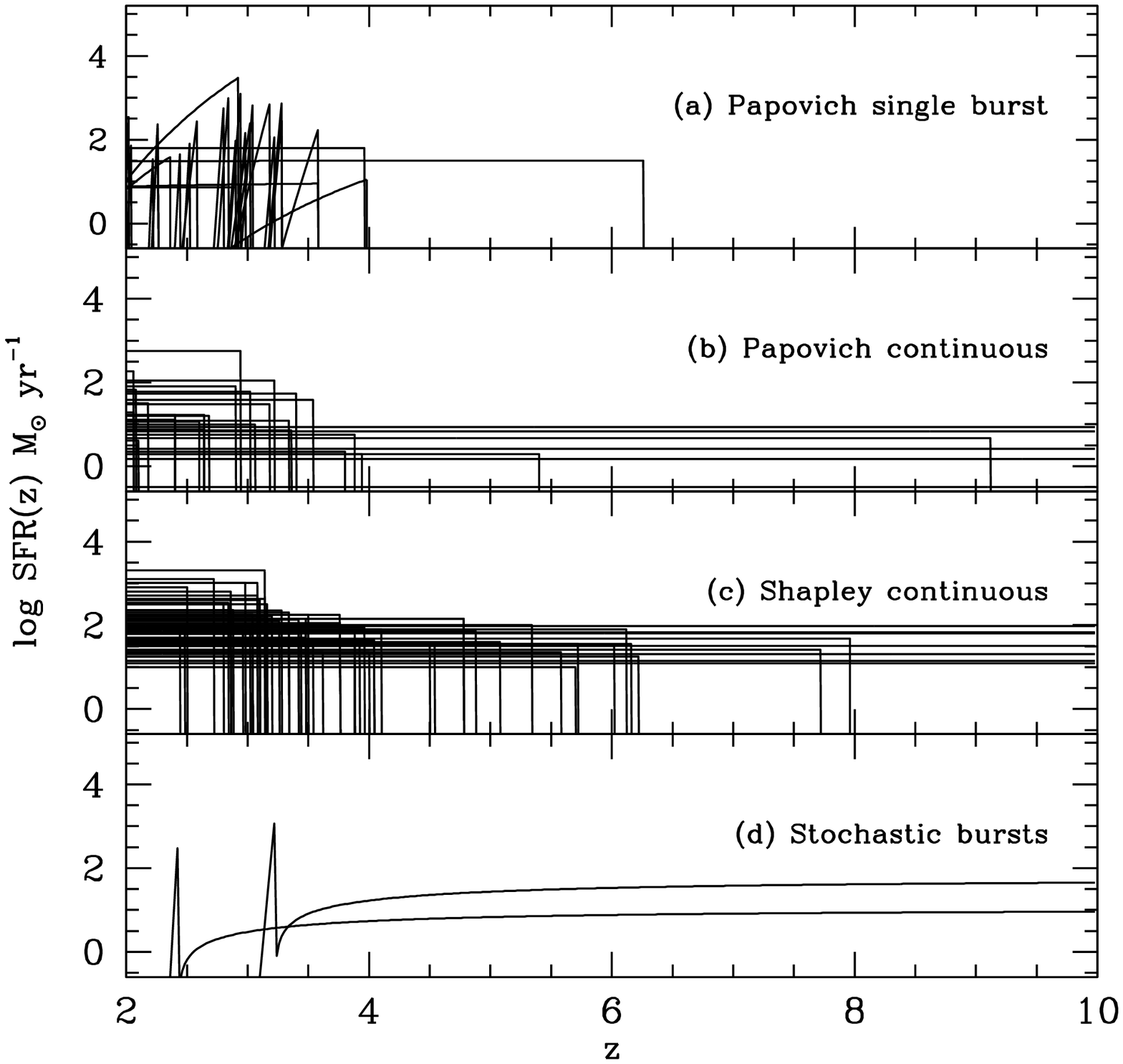}}
\caption{\protect{\label{fig1}}
Star-formation rate vs. time for individual galaxies, as inferred from
the SED models. The top panel shows the best-fit models of
type ($i$) described in the text from \citet{PDF01}. Panel (b) shows the
star-formation histories from models of type ($ii$) characterized by a 
stellar mass $M$ and an age, with a constant star-formation rate once
the galaxy has formed. The models shown here are the oldest ones
consistent with the SEDs within the 95\% confidence interval.
Panel (c) shows models with continuous star-formation
using the best-fit parameters for the \citet{SSADGP01} sample. 
Panel (d) shows two examples of the stochastic burst model described 
in the text applied to galaxies 97 and 1115 in the PDF01 sample. 
}
\end{figure}

\begin{figure}
\centerline{\plotone{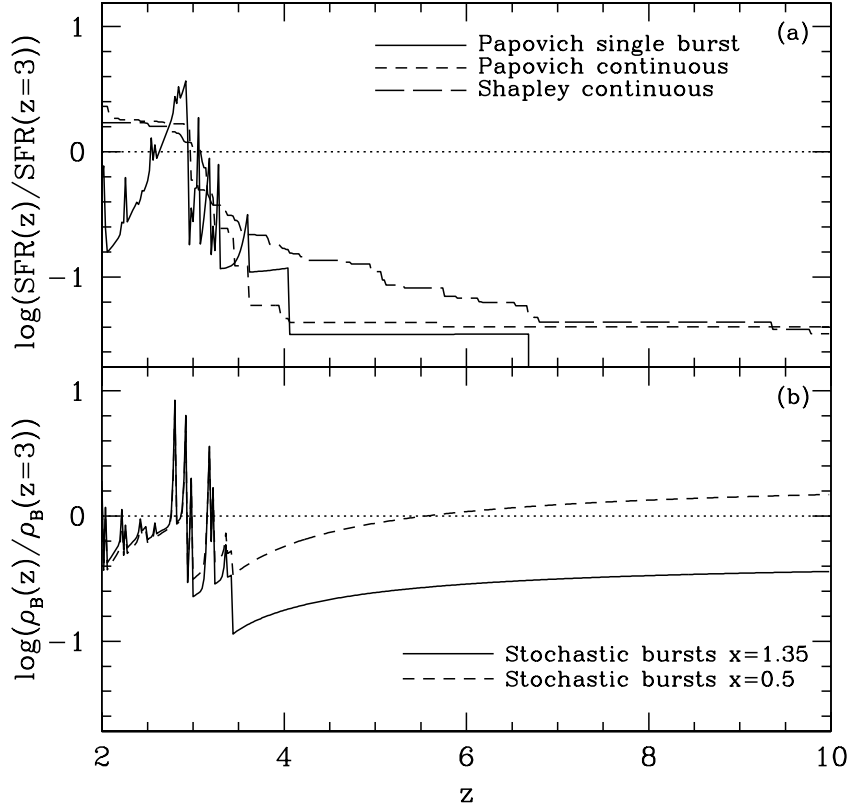}}
\caption{\protect{\label{fig2}}
Top panel --- star-formation density vs. time 
for the monotonic models, normalized
to the mean in the range at $2.5 < z < 3.5$. The solid curve is for the
PDF01 single-burst models.  The short-dashed curve is for their continous
star-formation models. The long-dashed curve is for the \citet{SSADGP01} continuous star-formation models. Bottom  panel ---
rest-frame B-band luminosity density vs. time for the
stochastic burst models with a Salpeter IMF (solid) and a top-heavy
IMF with $x=0.5$ (dashed).  For fixed IMF in the stochastic-burst model
the B-band luminosity density roughly scales with the star-formation
rate. As the IMF is varied the zeropoint of this scaling changes, so it
is more relevant to consider luminosity densities.  The B-band is shown
because that is what we calculate from the power-law fading model. The
UV luminosity density is more relevant for the discussion of
reionization and the $z=4$ luminosity function. Even for an extreme IMF
slope of $x=0.3$, the $m(1500{\rm \AA})-B$ colors and $m(860{\rm \AA})-B$ colors are
within 0.2  and 0.3 mag, respectively, of the colors for the Salpeter
IMF.
}
\end{figure}

\fi

\end{document}